\documentclass[preprint,showpacs,amsmath,amssymb]{revtex4}
\usepackage{graphicx}
\usepackage{amsfonts}
\usepackage{dcolumn}
\usepackage{bm}

\usepackage{CJK}

\begin{document}
\title{Double phase transition of the Ising model in core-periphery networks}

\author{Hanshuang Chen$^{1}$}\email{chenhshf@ahu.edu.cn}

\author{Haifeng Zhang$^2$}

\author{Chuansheng Shen$^{3}$}

\affiliation{$^1$School of Physics and Materials Science, Anhui
University, Hefei, 230601, China \\$^2$School of Mathematical
Science, Anhui University, Hefei, 230601, China \\ $^3$Department of
Physics, Anqing Normal University, Anqing, 246011, China}

\date{\today}

\begin{abstract}
We study the phase transition of the Ising model in networks with
core-periphery structures. By Monte Carlo simulations, we show that
prior to the order-disorder phase transition the system organizes
into an inhomogeneous intermediate phase in which core nodes are
much more ordered than peripheral nodes. Interestingly, the
susceptibility shows double peaks at two distinct temperatures. We
find that, if the connections between core and periphery increase
linearly with network size, the first peak does not exhibit any
size-dependent effect, and the second one diverges in the limit of
infinite network size. Otherwise, if the connections between core
and periphery scale sub-linearly with the network size, both peaks
of the susceptibility diverge as power laws in the thermodynamic
limit. This suggests the appearance of a double transition
phenomenon in the Ising model for the latter case. Moreover, we
develop a mean-field theory that agrees well with the simulations.

\end{abstract}
 \maketitle

\section{Introduction}
Phase transitions and critical phenomena on complex networks have
been a subject of intense research in statistical physics and many
other disciplines
\cite{PRP06000175,PRP08000093,RMP08001275,PRP2014}. Contrary to
regular lattices in the Euclidean space, complex networks are
usually characterized by a highly heterogeneous connectivity among
nodes, such as power-law degree distributions \cite{NewmanBook}.
Owing to the heterogeneity, phase transitions and critical phenomena
on complex networks are drastically different from those on regular
lattices. Examples range from the anomalous behavior of Ising model
\cite{PHA02000260,PLA02000166,PRE02016104,EPB02000191,PhysRevLett.104.218701}
to a vanishing percolation threshold
\cite{PhysRevLett.85.4626,PRL00005468} and the absence of epidemic
threshold that separates healthy and endemic phases
\cite{PRL01003200,PhysRevLett.90.028701,PhysRevLett.111.068701} as
well as explosive emergence of phase transitions
\cite{PhysRevLett.106.128701,PRP2016(2)}. On the other hand, many
real-world networks exhibit a typical mesoscopic structure such as
community structure \cite{PhysRep10000075,PNAS06008577}. A community
is a group of nodes that are densely interconnected and sparsely
connected to nodes in different communities. Such a community
structure is also one of essential ingredients for determining
dynamics on complex networks
\cite{PNAS2008.105.1118,PhysRevLett.96.114102}. In particular, it
was shown that in equilibrium
\cite{PhysRevE.80.025101,PhysRevE.83.046124} and nonequilibrium
\cite{CPL2015} Ising models community structure can lead to a novel
metastable phase in which spin orientation aligns with those in the
same community but disaligns with those in different communities.

Core-periphery structure is another mesoscale structure of networks,
with which a network is consisted of two groups of nodes called the
core and periphery. Core nodes are densely interconnected,
peripheral nodes are connected to core nodes to different extents,
and peripheral nodes are sparsely interconnected
\cite{Borgatti2000,J.ComplexNetw.2003,PhysRevE.72.046111,J.Stat.Mech.2016.023401,PhysRevE.93.022306,SIAMRev.2017}.
Core-periphery structure has been found in various networks,
including brain networks \cite{PLOSComput.Biol.2013}, protein
interaction networks \cite{Algo.Mol.Biol.2015}, social networks
\cite{Borgatti2000,SIAMJ.Appl.Math.2014}, transportation networks
\cite{PhysRevE.72.046111,PhysRevE.89.032810}, and so forth. Since
Borgatti and Everett \cite{Borgatti2000} introduced the first
quantitative formulation of core-periphery structure, many
algorithms have been developed for detecting the core-periphery
structure
\cite{Borgatti2000,Boyd2010,PhysRevE.72.046111,PhysRevE.89.032810,SIAMJ.Appl.Math.2014,PhysRevE.96.052313,arXivMasuda,PhysRevE.91.032803,Xiang2018Chaos}.
However, little attention has been paid to the dynamics on networks
with core-periphery structure. Recently, Verma \emph{et al.}
\cite{NCVerma2016} proposed a simple pruning process based on
removal of underutilized links and redistribution of loads and found
that such a process is responsible for the emergence of
core-periphery structure.

In the present work, we aim to study how would the core-periphery
structure impact the phase transition of Ising model. By Monte Carlo
(MC) simulation and a mean-field analysis, we show that an
intermediate phase emerges when the temperature is lower than the
critical one. Such an intermediate phase is rather inhomogeneous.
That is, core nodes are much more ordered than peripheral nodes. We
also find that the susceptibility exhibits a double-peak profile as
the temperature varies. We show that, on the one hand, if the number
of the connections between core and periphery is linear with the
network size, the height of the first peak is finite and does not
have a size-dependent effect, while the second one diverges in the
thermodynamic limit. On the other hand, if the connections between
core and periphery increase sub-linearly with the network size, both
peaks of the susceptibility diverge as power laws, which indicates
the occurrence of a double phase transition in the Ising model on
the disordered network systems. We should note that the
double-peaked phenomenon in susceptibility was reported recently in
percolation models \cite{PhysRevX.4.041020,PhysRevX.6.021002} and in
epidemic spreading models
\cite{PhysRevE.91.012816,PNAS2017.114.8969}.

\section{Model and Method}
We consider the Ising model on a network whose Hamiltonian is given
by,
\begin{equation}
\mathcal {H} =  - J\sum\limits_{i < j} {{A_{ij}}{\sigma _i}} {\sigma
_j} - h\sum\limits_i {{\sigma _i}},\label{eq1}
\end{equation}
where $\sigma_i \in \{  + 1, - 1\}$ is the spin variable of node
$i$, $J>0$ is the ferromagnetic interaction constant, and $h$ is the
external magnetic field. The network is described by an adjacency
matrix whose elements $A_{ij}$ are defined as $A_{ij}=A_{ji}=1$ if
nodes $i$ and $j$ are connected, and zero otherwise.

The network consists of $N$ nodes and $M=N\left\langle k
\right\rangle/2$ undirected edges, where $\left\langle k
\right\rangle$ is the average degree of the network. We pick a
fraction $p_c$ of nodes as core nodes, and the remaining $1-p_c$
fraction of nodes as peripheral nodes. We introduce the parameters
$\pi_{cc}$, $\pi_{cp}$, and $\pi_{pp}$ as the connectivity
probabilities among nodes in core-core, core-periphery, and
periphery-periphery, respectively. The number of edges in the
network can be computed by
\begin{equation}
N \left\langle k \right\rangle /2= \pi_{cc}N_c(N_c-1)/2+\pi_{cp}N_c
N_p+\pi_{pp}N_p(N_p-1)/2,\label{eq2}
\end{equation}
where $N_c=p_cN$ and $N_p=(1-p_c)N$ are the number of core nodes and
peripheral nodes, respectively. Assuming that $N_c,N_p\gg1$, Eq.
(\ref{eq2}) can be rewritten as
\begin{equation}
\bar \pi= \pi_{cc} p_c^2+2 \pi_{cp} p_c(1-p_c)+ \pi_{pp}(1-p_c)^2,
\label{eq3}
\end{equation}
where $\bar \pi=\left\langle k \right\rangle/(N-1)$ is the average
probability that each node is connected to the other nodes. By
defining $r_1=\pi_{cp}/\pi_{cc}$ and $r_2=\pi_{pp}/\pi_{cc}$,
$\pi_{cc}$ is thus expressed as
\begin{equation}
\pi_{cc}=\bar \pi/(p_c^2+2r_1 p_c(1-p_c)+r_2(1-p_c)^2), \label{eq4}
\end{equation}
and $\pi_{cp}=r_1 \pi_{cc}$, $\pi_{pp}=r_2 \pi_{cc}$. If
$r_1=r_2=1$, the resulting networks are Erd\"os-R\'enyi random
graphs. If $r_1 \sim \mathcal {O}(1)$ and $r_2 \sim \mathcal
{O}(0)$, the resulting networks have the characteristics of
core-periphery structure. The main aim of the present work is to
study the phase transition behaviors of Ising model on networks with
core-periphery structure.

We perform MC simulation with the Glauber dynamics. At each
elementary step, one node is randomly chosen and try to flip its
spin with the probability $1/\left(1+\exp(\beta \Delta E)\right)$,
where $\beta=1/\left(k_B T \right)$ is the inverse temperature,
$k_B$ is the Boltzmann constant, and $\Delta E$ is the change of the
system's energy due to the flipping trial. On each MC step (MCS),
each node is tried to update its spin once on average. To
characterize the phase behavior of the network, we need to define
three magnetizations: the average magnetization $m
=N^{-1}\sum\nolimits_{i = 1}^N {\sigma_i}$ of all the nodes, the
average magnetization $m_c= N_c^{-1}\sum\nolimits_{i \in \mathcal
{C}}{\sigma _i}$ of all the core nodes, and the average
magnetization $m_p= N_p^{-1}\sum\nolimits_{i \in \mathcal
{P}}{\sigma _i}$ of all the peripheral nodes, where $\mathcal {C}$
and $\mathcal {P}$ denote the sets of core nodes and peripheral
nodes, respectively. To make the system in equilibrium, the first
$10^5$ MCS are discarded and the following $10^5$ MCS are used to
calculate ensemble averages of the physical quantities. At the
critical region, larger runs are performed with $2\times10^5$ MCS to
reach the steady state and $10^6$ for computing the averages.

\section{Results}
\begin{figure*}
\centerline{\includegraphics*[width=1.0\columnwidth]{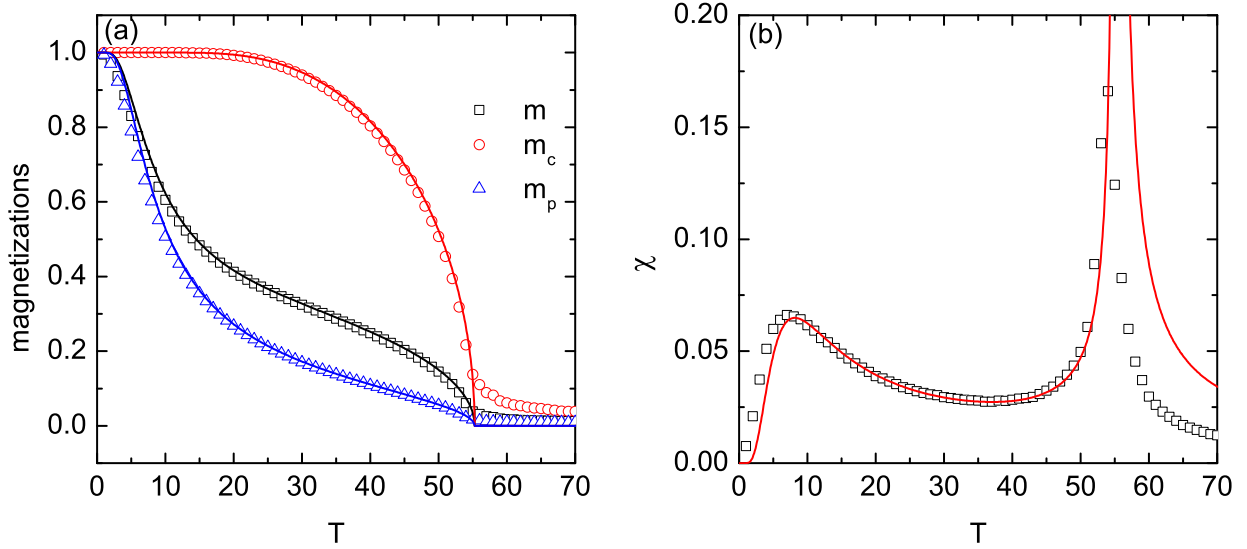}}
\caption{(color online) Phase transition in zero field $h=0$. (a)
The magnetizations $m$, $m_c$ and $m_p$ as functions of the
temperature $T$ (in unit of $J/k_B$). (b) The susceptibility $\chi$
as a function of $T$. The network parameters are $N=10000$,
$\left\langle k \right\rangle=20$, $p_c=0.2$, $r_1=0.1$, and
$r_2=0.005$. Symbols and lines indicate the MC simulation results
and theoretical ones, respectively. \label{fig1}}
\end{figure*}

Firstly, we demonstrate the results on the network with $N=10000$,
$\left\langle k \right\rangle=20$, $p_c=0.2$, $r_1=0.1$, and
$r_2=0.005$. Obviously, the network has a core-periphery structure.
Fig. \ref{fig1}(a) shows $m$, $m_c$, and $m_p$ as functions of the
temperature $T$ in the absence of external field, namely $h=0$. As
$T$ increases from zero, $m_p$ decreases much more quickly than
$m_c$. If $T$ is larger than a critical value $T_c$, both $m_p$ and
$m_c$ approach zero and a disordered paramagnetic phase emerges. For
$T$ between zero and $T_c$, there exists an intermediate phase in
which core nodes are much ordered than peripheral nodes. Such an
intermediate phase is caused by the core-periphery structure of the
network where the connectivity between core nodes is much denser
than that between peripheral nodes. Fig. \ref{fig1}(b) shows the
susceptibility $\chi$ as a function of $T$. Here $\chi$ is
calculated by the fluctuation of the magnetization $m$ according to
fluctuation-dissipation theorem, $\chi = \beta N \left[
{\left\langle {{m^2}} \right\rangle  - {{\left\langle m
\right\rangle }^2}}\right]$, where $\left\langle \cdot
\right\rangle$ denotes the averages taken in the stationary regime.
Interestingly, $\chi$ exhibits double peaks at two different $T$,
$T_{c_1}$ and $T_{c_2}$ with $T_{c_1}<T_{c_2}$, which seems to
indicate the existence of a double phase transition.

\begin{figure*}
\centerline{\includegraphics*[width=1.0\columnwidth]{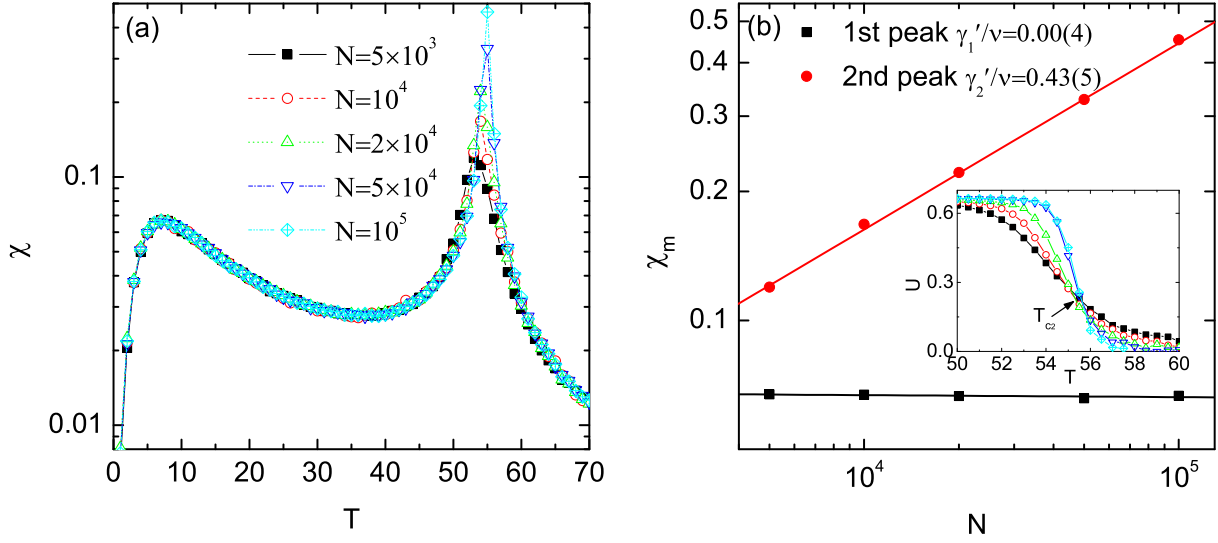}}
\caption{(color online) (a) The susceptibility $\chi$ as a function
of the temperature $T$ for different network size $N$. (b) The
maximal susceptibility $\chi_m$ as functions of $N$ with a double
logarithmic coordinate. The other parameters are the same as those
in Fig. \ref{fig1}. The lines in (b) indicate the linear fittings.
The inset in Fig.\ref{fig2}(b) shows the fourth-order cumulant $U$
as a function of $T$ for different $N$. $U(N)$ intercept each other
at the critical temperature $T_{c_2}$. \label{fig2}}
\end{figure*}

Since phase transition actually happens in the thermodynamic limit,
we consider the size effect of $\chi$ as follows. In Fig.
\ref{fig2}(a), we show $\chi$ as a function of $T$ for several
different $N$. One can see that the location of the first peak does
not change with the network size $N$ and its height $\chi_{m_1}$
does not change with $N$ either. However, unlike the first peak of
$\chi$, the location of the second peak shifts to a larger
temperature and its height $\chi_{m_2}$ increases as $N$ increases.
In Fig. \ref{fig2}(b), we show that how $\chi_{m_1}$ and
$\chi_{m_2}$ vary with $N$. In a double logarithmic coordinate,
$\chi_{m_1}$ and $\chi_{m_2}$ can be well fitted linearly, i.e,
$\chi_{m_1}\sim N^{\gamma^{\prime}_1/\nu}$ and $\chi_{m_2}\sim
N^{\gamma^\prime_2/\nu}$, with the exponents $\gamma^\prime_1/\nu
\simeq 0$ and $\gamma^\prime_2/\nu = 0.43(5)$. This suggests that
only $\chi_{m_2}$ diverges in the limit of $N\rightarrow \infty$. We
call the phenomenon a \emph{pseudo}-double phase transition. The
singularity in $\chi_{m_2}$ indicates an actual phase transition
will occur at a certain temperature $T_{c_2}$. To determine
$T_{c_2}$, we calculate the Binder's fourth-order cumulant, defined
as $U= 1 - {{\left\langle {{m^4}} \right\rangle } \mathord{\left/
 {\vphantom {{\left\langle {{m^4}} \right\rangle } {\left[ {3{{\left\langle {{m^2}} \right\rangle }^2}} \right]}}} \right.
 \kern-\nulldelimiterspace} {\left[ {3{{\left\langle {{m^2}} \right\rangle }^2}}
 \right]}}$. $T_{c_2}$ is determined as the point where the curves $U \sim T$ for different
 $N$ intercept each other. From the inset of Fig. \ref{fig2}(b), we estimate
 $T_{c_2}\simeq 55.5$.

To proceed a theoretical analysis, let us start with the mean-field
equations for $m_c$ and $m_p$, given by
\cite{PLA02000166,Chen2013JSM}

\begin{subequations} \label{eq5}
\begin{equation}
{m_c} = \tanh \left[ {\beta \left( {{k_{cc}}{m_c} + {k_{cp}}{m_p}}
\right) + \beta h} \right],
\end{equation}
\begin{equation}
{m_p} = \tanh \left[ {\beta \left( {{k_{pc}}{m_c} + {k_{pp}}{m_p}}
\right) + \beta h} \right],
\end{equation}
\end{subequations}
where $k_{cc}=\pi_{cc}(N_c-1)$ and $k_{cp}=\pi_{cp} N_p$ are the
connectivity numbers of a core node to other core nodes and
peripheral nodes, respectively. Likewise, $k_{pc}=\pi_{pc}N_c$ and
$k_{pp}=\pi_{pp}(N_p-1)$ are the connectivity numbers of a
peripheral node to core nodes and other peripheral nodes,
respectively.

For $h=0$, one notices that $m_c=m_p=0$ is always a set of solutions
of Eq. (\ref{eq5}). This set of trivial solution corresponds to the
paramagnetic phase. To determine the stability of the trivial
solution, we linearize Eq. (\ref{eq5}) around $m_c=m_p=0$, yielding
\begin{equation}
\hat{\textbf{m}}=\textbf{J} \hat{\textbf{m}}. \label{eq6}
\end{equation}
Here $\hat{ \textbf{{m}}}=(m_c, m_p)^\top$ with $\top$ denoting the
transpose, and
\begin{equation}
\textbf{J} = \beta \left( {\begin{array}{*{20}{c}}
  {{k_{cc}}}&{{k_{cp}}} \\
  {{k_{pc}}}&{{k_{pp}}}
\end{array}} \right) \label{eq7}
\end{equation}
is Jacobian matrix. The nonzero solutions of $\hat{\textbf{m}}$
exist when the leading eigenvalue of $\textbf{J}$ is less than one,
yielding the critical temperature,
\begin{equation}
{T_{c_2}} = \frac{{2\left( {{k_{cc}}{k_{pp}} - {k_{cp}}{k_{pc}}}
\right)}}{{{k_{cc}} + {k_{pp}} - \sqrt {{{\left( {{k_{cc}} -
{k_{pp}}} \right)}^2} + 4{k_{cp}}{k_{pc}}} }}. \label{eq8}
\end{equation}

Since the susceptibility is defined as
\begin{equation}
\chi (T,h) = {\left( {\frac{{\partial m}}{{\partial h}}}
\right)_T},\label{eq9}
\end{equation}
we take the partial derivation with respect to $h$ for Eq.
(\ref{eq5}), one has

\begin{subequations}\label{eq10}
\begin{equation}
  {\chi _c} = \beta \left( {1 - m_c^2} \right)\left( {{k_{cc}}{\chi _c} + {k_{cp}}{\chi _p}}+1
  \right),
\end{equation}
\begin{equation}
   {\chi _p} = \beta \left( {1 - m_p^2} \right)\left( {{k_{pc}}{\chi _c} + {k_{pp}}{\chi _p}}+1
   \right).
\end{equation}
\end{subequations}

Solving the above equations, one obtains
\begin{widetext}
\begin{subequations}\label{eq11}
\begin{equation}
  {\chi _c} = \frac{{\beta \left( {1 - m_c^2} \right)\left[ {\beta \left( {1 - m_p^2} \right)\left( {{k_{cp}} - {k_{pp}}} \right) + 1} \right]}}{{\left( {{k_{cc}}{k_{pp}} - {k_{cp}}{k_{pc}}} \right){\beta ^2}\left( {1 - m_c^2} \right)\left( {1 - m_p^2} \right) - \left[ {{k_{cc}}\left( {1 - m_c^2} \right) + {k_{pp}}\left( {1 - m_p^2} \right)} \right]\beta  +
  1}},
\end{equation}
\begin{equation}
{\chi _p} = \frac{{\beta \left( {1 - m_p^2} \right)\left[ {\beta
\left( {1 - m_c^2} \right)\left( {{k_{pc}} - {k_{cc}}} \right) + 1}
\right]}}{{\left( {{k_{cc}}{k_{pp}} - {k_{cp}}{k_{pc}}}
\right){\beta ^2}\left( {1 - m_c^2} \right)\left( {1 - m_p^2}
\right) - \left[ {{k_{cc}}\left( {1 - m_c^2} \right) +
{k_{pp}}\left( {1 - m_p^2} \right)} \right]\beta  + 1}}.
\end{equation}
\end{subequations}
\end{widetext}
At $T=T_{c_2}$, $m_c=m_p=0$ and the denominators on the right hand
side of Eq. (\ref{eq11}) equal to zero, such that the
susceptibilities $\chi_c$ and $\chi_p$ diverge at $T=T_{c_2}$. For
$T\neq T_{c_2}$, one can numerically solve Eq. (\ref{eq5}) to obtain
$m_c$ and $m_p$, as well as $m=p_c m_c+(1-p_c) m_p$. Furthermore,
substituting $m_c$ and $m_p$ into Eq. (\ref{eq11}) one comes to
$\chi=p_c \chi_c+(1-p_c) \chi_p$. As shown in Fig. \ref{fig1}(b) by
the lines, the theory gives that the first peak of $\chi$ occurs at
$T_{c_1}=8.2$ and $\chi$ diverges at $T_{c_2}=55.3$ that agree well
with the MC simulations.

\begin{figure}
\centerline{\includegraphics*[width=1.0\columnwidth]{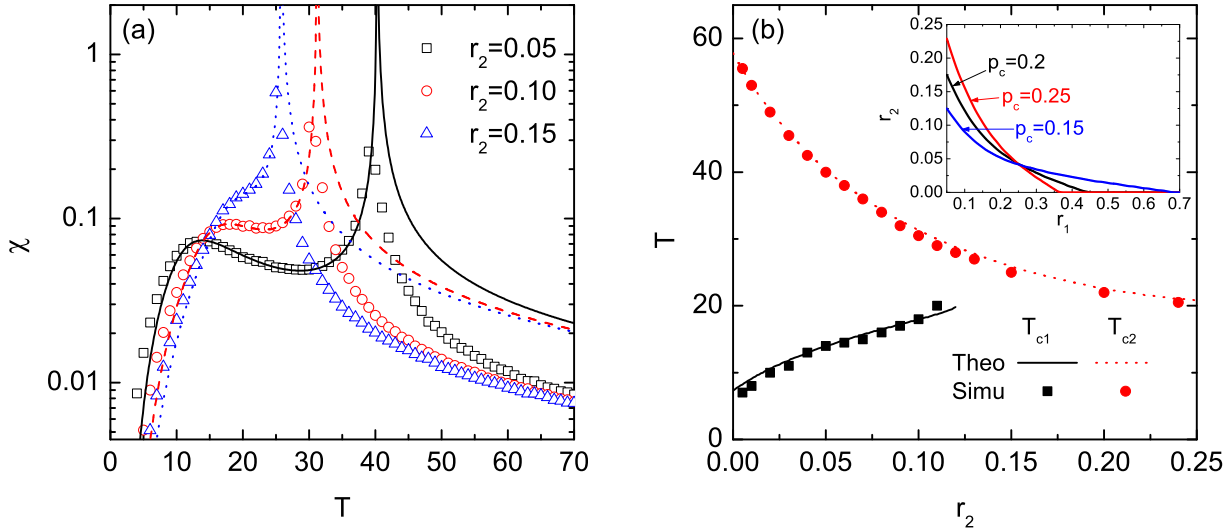}}
\caption{(color online) (a) The susceptibility $\chi$ as a function
of the temperature $T$ for a fixed $r_1=0.1$ and three different
$r_2$. (b) The peaked temperatures, $T_{c_1}$ and $T_{c_2}$, as
functions of $r_2$ for a fixed $r_1=0.1$. The other parameters are
the same as those in Fig. \ref{fig1}. Symbols and lines indicate the
MC simulation results and theoretical ones, respectively. The inset
in (b) shows that the threshold value $\tilde r_2$ as a function of
$r_1$ for three different $p_c$. As $T$ varies, $\chi$ has two peaks
in the region below the curve $\tilde r_2\sim r_1$, and has one
single peak above the curve. \label{fig3}}
\end{figure}

To investigate the effect of core-periphery structure on the phase
transition, we show the results for three distinct $r_2$ but with a
fixed $r_1=0.1$, as shown in Fig. \ref{fig3}(a). The larger the
value of $r_2$ is, the weaker core-periphery structure the network
has. One can see that for $r_2=0.05$, the pseudo-double phase
transition is still observed. For a lager $r_2=0.1$, the double-peak
phenomenon is not obvious. However, for $r_2=0.15$ $\chi$ exhibits
only one peak as usual. This implies that there exists a threshold
value of $r_2$ above which the pseudo-double phase transition
phenomenon is destroyed. In Fig. \ref{fig3}(b), we show the two
peaked temperatures, $T_{c_1}$ and $T_{c_2}$, as functions of $r_2$
with the fixed $r_1=0.1$. The first peaked-temperature $T_{c_1}$
increases with $r_2$, and terminates at the threshold value of
$r_2=\tilde r_{2}$. Our theory predicts $\tilde r_2=0.122$ that is
very close to the simulation value of $\tilde r_2=0.11$. The second
peaked-temperature $T_{c_2}$ decreases with $r_2$ and
asymptomatically approaches the average degree $\left\langle k
\right\rangle=20$ as $r_2\rightarrow 1$. In the inset of Fig.
\ref{fig3}(b), we show that as $r_2$ increases the threshold value
$\tilde r_2$ is decreased monotonically, and vanishes for
$r_1>0.44$. This implies that for $r_1>0.44$ there is no double
peaks in $\chi$ no matter what the value of $r_2$ is. Furthermore,
we show the fraction $p_c$ of core nodes has an impact on $\tilde
r_2$, as drawn three different $p_c$ in the inset of Fig.
\ref{fig3}(b).

\begin{figure}
\centerline{\includegraphics*[width=1.0\columnwidth]{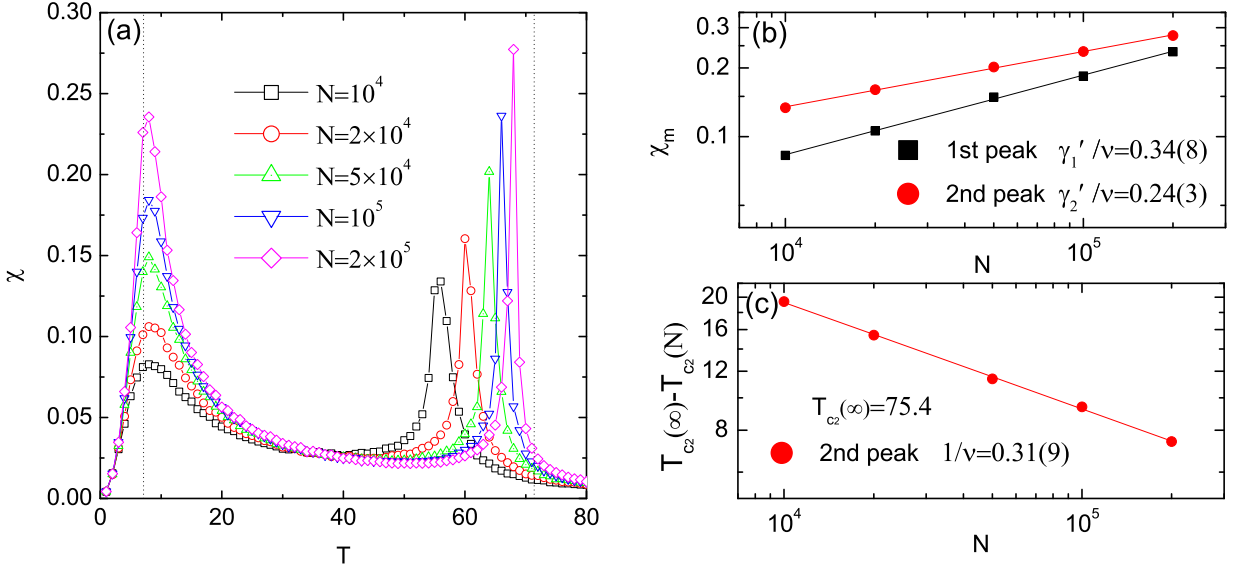}}
\caption{(color online) (a) The susceptibility $\chi$ as a function
of the temperature $T$ in core-periphery networks whose number of
connections between core nodes and periphery nodes decays
sub-linearly with the network size $N$, i.e., as $N^{1-\alpha}$ with
$\alpha=0.5$. The other networks parameters: the average degree
$\left\langle k \right\rangle=20$, the fraction of core nodes
$p_c=0.2$, and the ratio of the average degrees of a core node to a
periphery node $r_d=k_c/k_p=10$ are fixed. The left and right
vertical dotted lines indicate two critical temperatures of
mean-field prediction, $T_{c_1}=k_p\simeq7.14$ and
$T_{c_2}=k_c\simeq71.4$, respectively. (b) The two peaked
susceptibilities, $\chi_{m_1}$ and $\chi_{m_2}$ as functions of $N$.
The solid lines correspond to the linear fittings in a double
logarithmic coordinate, as $\chi_{m_{1,2}} \sim
N^{\gamma_{1,2}^{\prime}/\nu}$. (c) The differences between
$T_{c_2}(\infty)$ and $T_{c_2}(N)$ scale with $N$ as $\Delta
T_{c_2}(N) \sim N^{-1/\nu}$. The solid line shows the linear
fitting. \label{fig4}}
\end{figure}

\begin{figure}
\centerline{\includegraphics*[width=1.0\columnwidth]{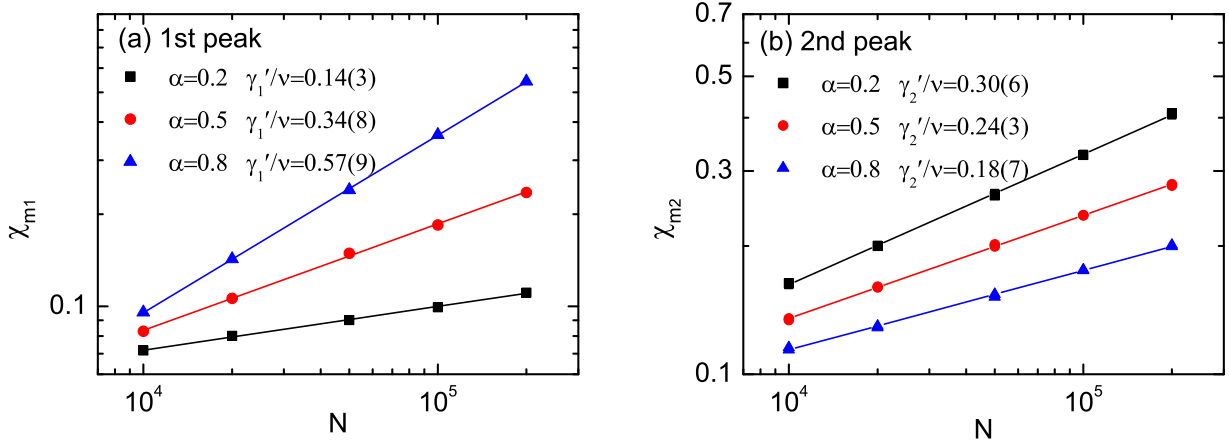}}
\caption{(color online)  The two peaked susceptibilities,
$\chi_{m_1}$ (a) and $\chi_{m_2}$ (b), as functions of $N$ for three
values of $\alpha=0.2$, 0.5, and 0.8. The solid lines correspond to
the linear fittings in a double logarithmic coordinate, as
$\chi_{m_{1,2}} \sim N^{\gamma_{1,2}^{\prime}/\nu}$. \label{fig5}}
\end{figure}

As shown in \cite{PhysRevX.4.041020} for the percolation model, a
\emph{true} double transition phenomenon is expected to occur if the
number of connections among nodes in the core and periphery scale
sub-linearly with the system size, i.e., as $N^{1-\alpha}$ with
$0<\alpha<1$. In this case, $k_{cp}$ and $k_{pc}$ become zero in the
thermodynamic limit, $m_c$ and $m_p$ are thus decoupled in
Eq.(\ref{eq5}) that allows for two distinct transition temperatures
in the spirit of mean-field theory, $T_{c_1}=k_{p}$ and
$T_{c_2}=k_{c}$, where $k_c$ and $k_p$ are the average degrees of a
core node and a periphery node, respectively. In our notation, this
is equivalent of making $\pi_{cp} \sim N^{-1-\alpha}$. Meanwhile, we
let the average degree $\left\langle k \right\rangle$ of the network
unchanged, and the ratio $r_d=k_c/k_p$ fixed. Thus, $k_c$ and $k_p$
can be obtained by the equality $\left\langle k
\right\rangle={k_c}{p_c} + {k_p}( {1 - {p_c}})$.  In
Fig.\ref{fig4}(a), we show $\chi$ as a function of $T$ for five
distinct $N$ with $\alpha=0.5$, $\left\langle k \right\rangle=20$,
$p_c=0.2$, and $r_d=10$. As expected, the susceptibility exhibits
two peaks whose maxima, $\chi_{m_1}$ and $\chi_{m_2}$, both increase
with $N$ in power-law ways $\chi_{m_{1,2}} \sim
N^{\gamma_{1,2}^{\prime}/\nu}$ (shown in Fig.\ref{fig4}(b)).
Therefore, both $\chi_{m_1}$ and $\chi_{m_2}$ diverge in the
thermodynamic limit such that a true double transition phenomenon
occurs in the networked Ising model with core-periphery structure.
On the other hand, the first peak is always located at $T=8$
regardless of the value of $N$. Such a position is very close to the
mean-field prediction $T_{c_1}=k_p\simeq 7.14$ (indicated by the
left vertical dotted line in Fig.\ref{fig4}(a)). The position
$T_{c_2}(N)$ of the second peak shifts to larger values of $T$ as
$N$ increases. In the limit of $N \rightarrow \infty$, $T_{c_2}(N)$
approaches an actual critical temperature
$T_{c_2}=T_{c_2}(N=\infty)$. As shown in Fig.\ref{fig4}(c), the
differences $\Delta T_{c_2}(N)=T_{c_2}-T_{c_2}(N)$ scale with $N$ as
$\Delta T_{c_2}(N) \sim N^{-1/\nu}$, with $1/\nu=0.319$ and
$T_{c_2}=75.4$ (approximately equals to the mean-field prediction
$T_{c_2}=k_c\simeq 71.4$, as indicated by the right vertical dotted
line in Fig.\ref{fig4}(a)). Finally, we consider the effect of
$\alpha$ on the critical exponents $\gamma^{\prime}_{1,2}/\nu$. In
Fig.\ref{fig5}, we show the power-law fits of $\chi_{m_{1,2}}$ as
$N$ for three distinct $\alpha$. One can see that  the power-law
exponent $\gamma^{\prime}_{1}/\nu$ is increased and
$\gamma^{\prime}_{2}/\nu$ is decreased as $\alpha$ increases.

\section{Conclusions}
In conclusion, we have studied the phase transition of the Ising
model in networks with core-periphery structure. We find that a
strong core-periphery structure can lead to the occurrence of an
intermediate phase prior to the order-disordered phase transition.
At the intermediate phase, the spin configuration in the network is
rather inhomogeneous. The core nodes are much more ordered than
peripheral nodes. Interestingly, the susceptibility peaks at two
distinct temperatures. We find that the susceptibility at the first
peaked temperature does show any size-dependent effect if the
connections between core and periphery are linear with the network
size $N$. Otherwise, if the connections between core and periphery
are sub-linear with $N$, the position of the first peaked
susceptibility does not shift with $N$ and its height diverges as
$N\rightarrow \infty$ in a power-law way. For the two cases, the
height of the second peak always increases with $N$ as a power law
and diverges in the limit of $N \rightarrow \infty$. The location of
the second peak increases with $N$ and asymptomatically approaches
the critical temperature $T_{c_2}$ of order-disorder phase
transition as $N \rightarrow \infty$. Therefore, the occurrence of a
double phase transition in the Ising model lies on the sub-linear
dependence of the connections between core and periphery on $N$,
which is consistent with the conclusion of \cite{PhysRevX.4.041020}.
Moreover, we develop a mean-field theory for calculating the
magnetization and susceptibility. The theory agrees well with the
simulations. In the future, it is expected that the phase transition
of other statistical physics models in networks with core-periphery
structure should be considered.

\begin{acknowledgments}
We acknowledge the supports from the National Natural Science
Foundation of China (Grants No. 11475003, No. 61473001), the Key
Scientific Research Fund of Anhui Provincial Education Department
(Grants No. KJ2016A015), ``211" Project of Anhui University (Grant
No. J01005106), and the Natural Science Foundation of Anhui Province
(Grant No. 1808085MF201).
\end{acknowledgments}


\end{document}